\documentclass[%
reprint,
superscriptaddress,
amsmath,amssymb,
aps,
pra,
floatfix,
]{revtex4-1}

\usepackage{graphicx}
\usepackage{dcolumn}
\usepackage{bm}
\usepackage{subfigure}
\usepackage{amsmath}
\usepackage{mathrsfs}
\usepackage{amssymb}
\usepackage{setspace}
\usepackage{array}
\usepackage{multirow}
\usepackage{float}
\usepackage{flushend}
\usepackage{footmisc}

\usepackage[pdfstartview=FitH,
CJKbookmarks=true,
colorlinks,
linkcolor=blue,
anchorcolor=blue,
citecolor=blue,
urlcolor=blue,
]{hyperref}

\renewcommand{\footnoterule}{
	\kern -4pt  
	\hrule width 0.18\linewidth height 0.6pt
	\kern 12pt 
}

\usepackage{braket}

\begin{document}

\preprint{APS/123-QED}

\title{Simulation of classical Ising-like magnetism with a Mott insulator of paired atoms}
\author{Ren Liao}
\email{liaoren@pku.edu.cn}
\affiliation{School of Electronics, Peking University, Beijing 100871, China}

\author{Jingxin Sun}%
\affiliation{School of Electronics, Peking University, Beijing 100871, China}

\author{Pengju Zhao}
\affiliation{School of Physics, Peking University, Beijing 100871, China}

\author{Shifeng Yang}
\affiliation{School of Electronics, Peking University, Beijing 100871, China}

\author{Hui Li}%
\affiliation{School of Electronics, Peking University, Beijing 100871, China}

\author{Xinyi Huang}%
\affiliation{School of Electronics, Peking University, Beijing 100871, China}

\author{Wei Xiong}%
\affiliation{School of Electronics, Peking University, Beijing 100871, China}

\author{Xiaoji Zhou}%
\affiliation{School of Electronics, Peking University, Beijing 100871, China}

\author{Dingping Li}
\affiliation{School of Physics, Peking University, Beijing 100871, China}

\author{Xiongjun Liu}
\affiliation{School of Physics, Peking University, Beijing 100871, China}

\author{Xuzong Chen}%
\email{xuzongchen@pku.edu.cn}
\affiliation{School of Electronics, Peking University, Beijing 100871, China}

\begin{abstract} 
Quantum simulation of the XXZ model with a two-component Bose or Fermi Hubbard model based on a Mott insulator background has been widely used in the investigations of quantum magnetism with ultracold neutral atoms. In most cases, the diagonal spin-spin interaction is always accompanied by a large spin-exchange interaction which hinders the formation of long-range magnetic order at low temperature. Here we show that the spin-exchange interaction can be strongly reduced in a Mott insulator of paired atoms, while the diagonal spin-spin interaction remains unaffected. Thus, the effective magnetic model is quite close to an exact classical Ising model in the textbook. And we analysed an experimentally achievable three-component Fermi-Hubbard model of $\mathrm{{}^{6}Li}$  with two hyperfine levels of atoms paired in the lattice. We find the long-range antiferromagnetic order of such a three-component Fermi-Hubbard model can be much stronger than that of a typical two-component Fermi-Hubbard model at low temperature. And we discussed the possiblity of simulating an exact two-dimensional ferromagnetic Ising model in a Mott insulator of paired bosonic atoms. Our results may be useful for experimental investigation of the long-range Ising-like magnetism with ultracold neutral atoms under thermal equilibrium.

\end{abstract}
\maketitle

\section{Introduction}


Quantum simulation \cite{Georgescu2014} of magnetic models has made lots of progress in recent years. In the systems of superconducting circuits \cite{Salathe2015,Gong2016,Barends2016, Harris2018,Alba2018} and trapped ions \cite{Friedenauer2008,Kim2010,Kim2011,Lanyon2011,Britton2012}, the simulated magnetic models are usually the Ising model with spatially dependent spin-spin interaction. And the number of simulated spins is usually limited due to the control difficulties. Another direction is to simulate a large-scale quantum magnet in an optical lattice by controlling the long-range dipole-dipole interaction of polar gases and Rydberg atoms. It is reported that the XXZ model \cite{Paz2013} and the antiferromagnetic Ising model \cite{Elmer2018,Labuhn2016,Schauss2018,Semeghini2021,Antoine2021} have been realized. And there are also many theoretical proposals of simulating a Heisenberg-like magnetic model with polar molecules \cite{Barnett2006,Gorshkov2011,Maria2011,Kaden2013} and theoretical and experimental works about simulating magnetic models with laser-dressed Rydberg atoms \cite{Peter2015,Zeiher2016,You2019,Simon2022}. However, the effective spin-spin interaction arised from dipole-dipole interaction is intrinsically long-range and usually inhomogeneous in the lattice. Moreover, it is typically hard to reach thermal equilibrium states in the lattice due to the complexity of dipole-dipole interactions.

Besides dipole-dipole interaction, the superexchange interaction between neutral atoms in an optical lattice is also widely used to simulate magnetic models. The effective spin-spin interaction derived from superexchange interaction is usually homogeneous throughout the lattice and only involves the nearest-neighbor spins. And thermal equilibrium can also be reached by atom collisions in most cases.  Apart from some special designs of magnetic models \cite{Greiner2011,Struck2011}, the effective magnetic models are described by a XXZ model with a nearest-neighbor spin-spin interaction $J_z\hat{S}_i^z\hat{S}_j^z$ and a spin-exchange interaction $J_{\perp}(\hat{S}_i^+\hat{S}_j^-+H.c.)$. And the antiferromagnetic spin correlations of various kinds of Fermi-Hubbard models \cite{Daniel2013,Russell2015,Bloch2016,Martin2016,Drewes2017,Mazurenko2017,Ozawa2018}, the spin-charge seperation of a hole-doped Fermi-Hubbard model \cite{Timon2017}, the propogation of magnons \cite{Ivana2020,Bloch2013} and spinons \cite{Bloch2020}  have been observed with ultracold neutral atoms. However, the spin-exchange interaction is usually quite large compared with the diagonal spin-spin interaction in these XXZ models realized with neutral atoms. Though there are reported studies of simulating an exact Ising model in a tilted lattice \cite{Greiner2011,Liao2021} and a tunable XYZ model with a $p$-band Mott insulator \cite{Pinheiro2013}, these methods suffer from the defect of a highly excited Mott insulator background which makes the effective magnetic models unstable in the lattice.

Here we propose a general method of simulating an Ising-like quantum magnet with the superexchange interaction of neutral atoms in an optical lattice. The main idea is to construct the low-energy effective magnetic models based on a Mott insulator of paired atoms. This idea is applicable to both bosonic atoms and fermionic atoms and has broad lattice compatability. Especially, an exact textbook Ising model with achievable thermal equilibrium can be simulated in certain cases.  The only requirement is to find suitable atom species to realize such kinds of Mott insulator without severe heating or loss caused by three-body recombination. These features make the paired-atom Mott insulator an ideal method to simulate the classical Ising-like models in an optical lattice.

\section{The superexchange interaction of paired atoms}

\begin{figure}[htpb]
	\includegraphics[width=\linewidth]{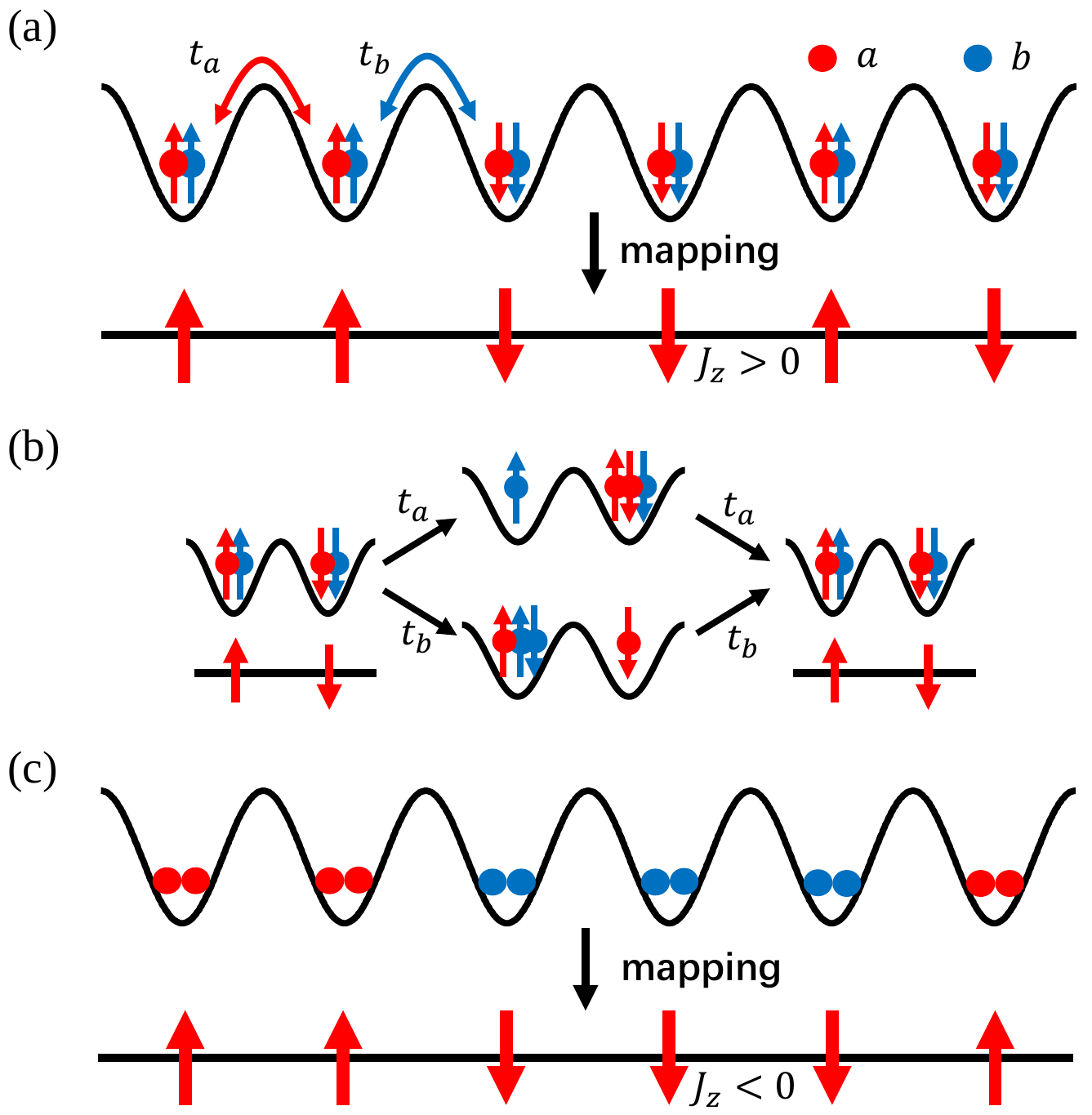}
	\caption{(a) The low-energy effective model of a Mott insulator with spin-paired fermionic atoms is an antiferromagnetic Ising model. At each site, the same spin components of each species of atoms are paired so that $|n_{i\uparrow}^a=1,n_{i\uparrow}^b=1\rangle$ and $|n_{i\downarrow}^a=1,n_{i\downarrow}^b=1\rangle$ can be mapped to spin up and down of a single spin, respectively. (b) The second-order low-energy superexchange interaction between spin-paired fermions. It can be seen that the spin-exchange terms are prohibited. (c) The Mott insulator of paired bosonic atoms generates an effective ferromagnetic Ising model when the on-site repulsive interaction of heteronuclear atom pairs $U^{ab}$ is larger than those of homonuclear atom pairs $U_a,U_b$. }
	\label{fig:pair_atoms}
\end{figure}

To begin with, we first analyse the superexchange interaction in a Mott insulator of paired atoms. In a two-species two-spin Fermi-Hubbard model shown in Fig. \ref{fig:pair_atoms}(a), the Hamiltonian can be written as
\begin{align}
\hat{H}_{pair}^{FH}=&\hat{H}_a^{FH}+\hat{H}_b^{FH}+\hat{H}_{ab}^{FH}, \label{eqn:H_pair}\\
\hat{H}_{s=a,b}^{FH}=&-t_s\sum_{\sigma=\uparrow\downarrow}\sum_{\langle ij\rangle} (\hat{s}_{i\sigma}^{\dagger}\hat{s}_{j\sigma}+H.c.)+ U_s\sum_i\hat{n}_{i\uparrow}^s\hat{n}_{i\downarrow}^s\notag \\
\hat{H}_{ab}^{FH}=&-U^{ab}\sum_{\sigma=\uparrow\downarrow}\sum_i  \hat{n}_{i\sigma}^a\hat{n}_{i\sigma}^b. \notag
\end{align}
Here $\langle ij\rangle$ represents the nearest-neighbor sites. And the parameters are set as $t_a,t_b\ll U_a,U_b,U^{ab}$. At the half-filling regime for each species of atoms, large $U_a,U_b$ requires each lattice site filled with only one a atom and one b atom. $U^{ab}$ provides a strong pairing interaction between the same spin components of a,b atoms so that there are only two possible occupation states $|\hat{n}_{i\uparrow}^a=1, \hat{n}_{i\uparrow}^b=1\rangle$ and $|\hat{n}_{i\downarrow}^a=1, \hat{n}_{i\downarrow}^b=1\rangle$ for the lowest-energy subspace, which can be mapped to  $|\hat{S}_i^z=1/2\rangle$ and $|\hat{S}_i^z=-1/2\rangle$ of a single $S=1/2$ spin, respectively. Up to third-order perturbations, the low-energy effective spin model of $\hat{H}_{pair}^{FH}$ in a cubic lattice can be written as 
\begin{align}
\hat{H}_{\mathrm{eff}}^{FH}=J_z^{FH}\sum_{\langle ij\rangle} \hat{S}_i^z\hat{S}_j^z
\end{align}
with $J_z^{FH}=\sum_{s=a,b}\frac{4t_s^2}{U^{ab}+U_s}$. Comparing with the typical two-component Fermi-Hubbard model of generating an low-energy effective Heisenberg model \cite{Efstratios1991}, we could find the spin pairing interaction $\hat{H}_{ab}$ is the key factor for the realization of an exact Ising model [Fig. \ref{fig:pair_atoms}(b)]. 

When we consider a Mott insulator of two species of bosonic atom pairs as shown in Fig. \ref{fig:pair_atoms}(c), the Hamiltonian of such a Bose-Hubbard model can be given as
\begin{align}
\hat{H}_{pair}^{BH}=&\hat{H}_a^{BH}+\hat{H}_b^{BH}+\hat{H}_{ab}^{BH}, \label{eqn:boson}\\
\hat{H}_{s=a,b}^{BH}=&-t_s\sum_{\langle ij\rangle} (\hat{s}_{i}^{\dagger}\hat{s}_{j}+H.c.) 
+\sum_i\frac{U_s}{2} \hat{n}_{is}(\hat{n}_{is}-1) \notag\\
\hat{H}_{ab}^{BH}=&U^{ab}\sum_i  \hat{n}_{ia}\hat{n}_{ib}.\notag
\end{align} 
Here $\hat{s}_{i\sigma}^{\dagger},\hat{s}_{i\sigma}$ ($s=a,b$) become bosonic operators. The  parameters satisfy $t_s\ll U_s,U^{ab}-U_s$ ($s=a,b$). With the same spin mapping in Fig. \ref{fig:pair_atoms}(c), the effective Hamiltonian becomes
\begin{align}
\hat{H}_{\mathrm{eff}}^{BH}= J_z^{BH}\sum_{\langle ij\rangle} \hat{S}_i^z\hat{S}_j^z
+h_z^{BH}\sum_i \hat{S}_i^z
\label{eqn:H_eff_BH}
\end{align}
with
\begin{align}
J_z^{BH}=\sum_{s=a,b}\left[\frac{4t_s^2}{2U^{ab}-U_s}-\frac{12t_s^2}{U_s}\right], \notag\\
h_z^{BH}=U_a-U_b+\frac{12t_b^2}{U_b}-\frac{12t_a^2}{U_a}. \notag
\end{align}
Here we igonored the boundary differences. It can be seen that $J_z^{BH}<0$ so that the ferromagnetic Ising model can also be exactly simulated. It is also noticeable that the magnitude of $|J_z^{BH}|$ can be serval times larger than a typical superexchange interaction with a magnitude $4t^2/U$. If such a model can be implemented in experiment, the requirement of a low spin temperature to observe the magnetic order induced by superexchange interaction can be strongly relieved.

\section{The three-component Fermi-Hubbard model} 
Eqn. (\ref{eqn:H_pair}) and (\ref{eqn:boson}) provide two ideal models of simulating an exact  Ising model with a Hubbard model. However, when considering experimental realization of $\hat{H}_{pair}^{FH}$ and $\hat{H}_{pair}^{BH}$, there are two nonnegligible concerns which need to be taken into account. One is the three-body recombination which may cause serious atom loss or heating in the lattice. The other is the realization of $U^{ab}$ should be in accordance with  $U_a$ and $U_b$, considering the on-site interactions are mainly tuned with Feshbach resonance by controlling the s-wave scattering length $a_{ss'}(B)$ ($s,s'$ are the same or two different atoms) with a magnetic field $B$ \cite{Cheng2010}. However, there are six different $a_{ss'}(B)$ curves between the four different atoms in Fig. \ref{fig:pair_atoms}(a) and three different $a_{ss'}(B)$ curves for the bosonic case, while the tuning parameter is only the magnetic field $B$. $\hat{H}_{pair}^{FH}$ is nearly impossible to be realized while $\hat{H}_{pair}^{BH}$ can be possibly realized with certain atom species. For example,   it is possible to simulate $\hat{H}_{pair}^{BH}$ exactly with the mixture of ${}^{87}\mathrm{Rb}$ $|F=1,m_F=-1\rangle$ (denoted as $a$ atoms) and ${}^{85}\mathrm{Rb}$ $|F=2,m_F=-2\rangle$ (denoted as $b$ atoms) considering $a_{aa}=100.4a_0$ \cite{Mertes2007}, $a_{bb}=-443a_0[1-10.7/(B-155.04)]$ \cite{Claussen2003} and $a_{ab}=213a_0[1-5.8/(B-265.4)]$ \cite{Papp2006}. Under a magnetic field of around $B=163.8$G, $a_{aa}=100.4a_0,a_{bb}=100.6a_0,a_{ab}=225.2a_0$ can be obtained. Here $a_0$ is the Bohr radius. Since $a_{aa}$ and $a_{ab}$ are quite insensitive to $B$ when $B<200$G, it is also very eaisly to tune $a_{aa}-a_{bb}$ at around $B=163\sim165$G.

\begin{figure}[tbp]
	\includegraphics[width=\linewidth]{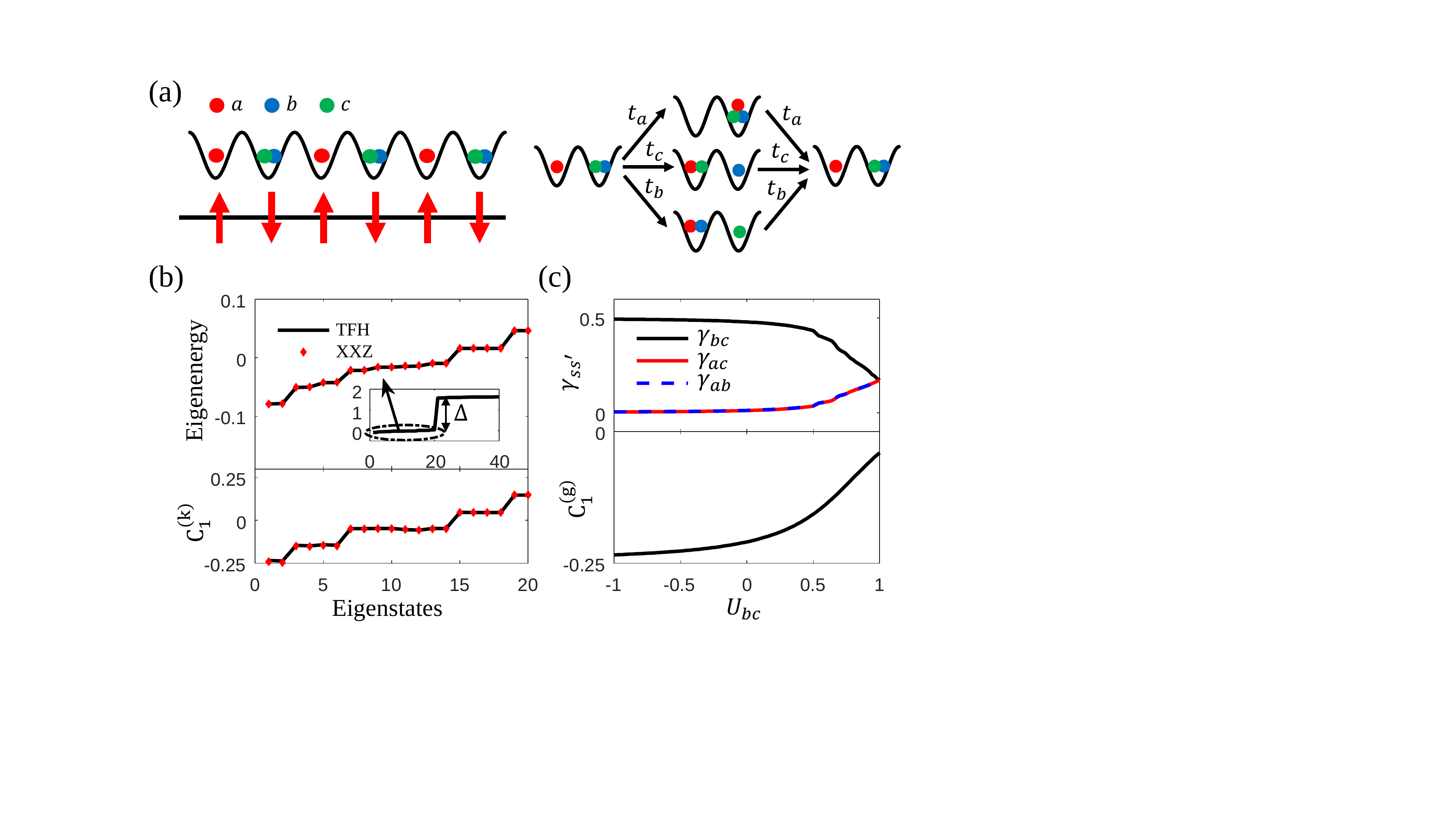}
	\caption{ The three-component Fermi-Hubbard model (TFH model). (a) Spin up and down can be represented by $a$ atoms and $bc$ atom pairs in each lattice site, respectively. (b) The eigenenergy spectrum and the nearest-neighbor spin correlations $C_1^{(k)}$ of the lowest $N_S=20$ eigenstates of the TFH model are both consistent with those of the  effective XXZ model. Here $k$ is the index of eigenstates. The parameters are set as $t_a=t_b=t_c=1/7, U_{ab}=U_{ac}=-U_{bc}=1,N_a=N_b=N_c=3,L=6$. (c) The effective nearest-neighbor spin correlation $C_1^{(g)}$ and the pairing ratios $\gamma_{ss'}$ ($ss'=ab,bc,ac$) of the ground state with respect to $U_{bc}$. When $U_{bc}$ is decreased from $U_{bc}=1$ to $U_{bc}=-1$, an antiferromagnetic ground state comes into existence gradually ($C_1^{\mathrm{(g)}}\rightarrow -1/4$) as b,c atoms begin to be paired ($\gamma_{bc}\rightarrow 0.5, \gamma_{ab}\rightarrow 0,\gamma_{ac}\rightarrow 0$).}
    \label{fig:theory}	
\end{figure}


\begin{figure}[thbp]
	\includegraphics[width=\linewidth]{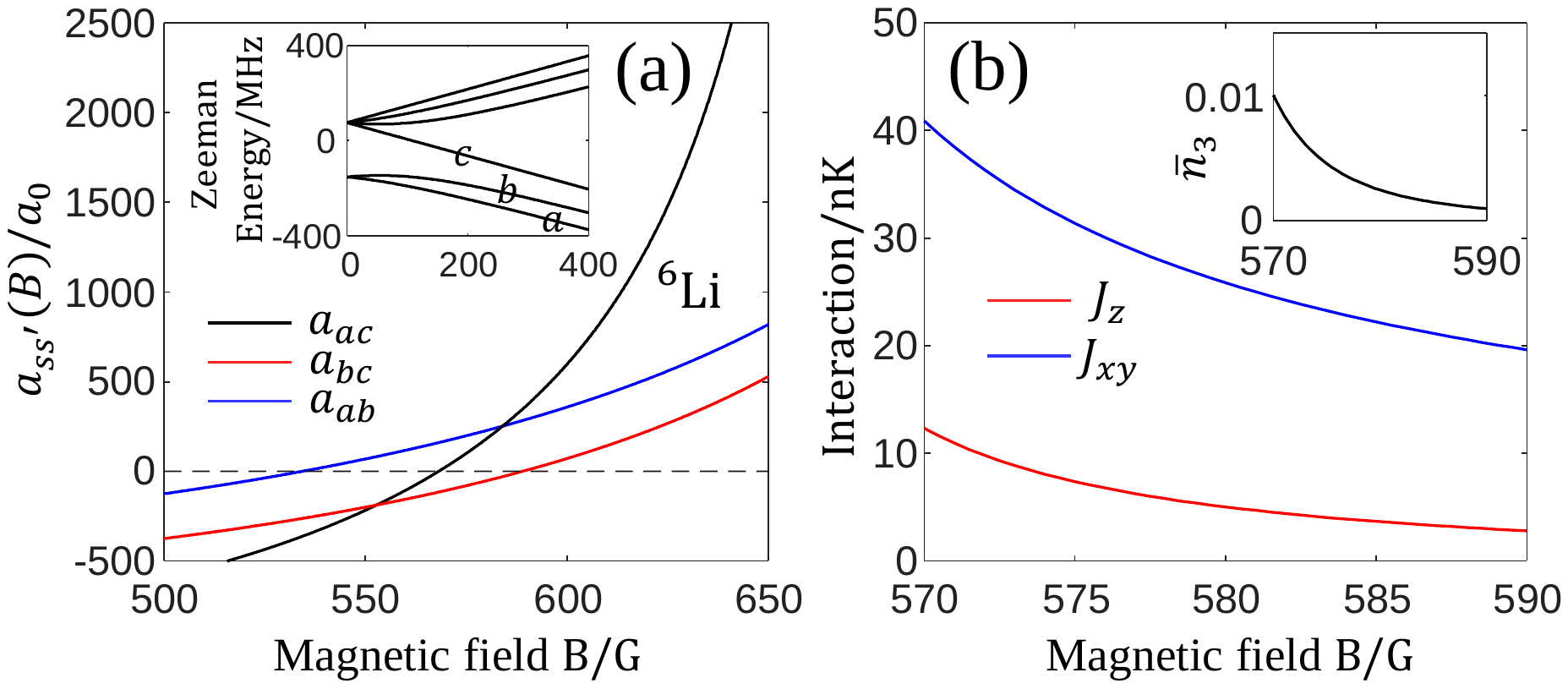}
	\caption{ Realization of the TFH model with ${}^{6}\mathrm{Li}$ atoms. (a) The lowest three hyperfine levels of ${}^6\mathrm{Li}$ can just be mapped to the $a,b,c$ atoms required in Fig. \ref{fig:theory}(a). The s-wave scattering lengths $a_{ss'}$ ($ss'=ab,ac,bc$) of ${}^6\mathrm{Li}$ satisfy $a_{ab},a_{ac}>0,a_{bc}<0$ when $570\mathrm{G}\leq B\leq590\mathrm{G}$. And the three-body recombination rate of a,b,c atoms reaches a minimum at around $B=560\sim 590 $G \cite{Ottenstein2008,Huckans2009}. (b) The estimated magnetic interaction $J_z,J_{xy}$ of the TFH model based on the experimental parameters of the Fermi-Hubbard model  realized with ${}^6\mathrm{Li}$ \cite{Mazurenko2017}. Here we assume $U_{ss'}(B)=a_{ss'}(B)/210a_0*6.50\mathrm{kHz}, t_a=t_b=t_c=0.90\mathrm{kHz}$.  The inlet is the mean three-body occupation $\bar{n}_3$ of the lowest $N_S$ eigenstates [FIG. \ref{fig:theory}(b)] per lattice site calculated in a 1D lattice with $L=6$. }
	\label{fig:parameter}
\end{figure}

Meanwhile, to indicate the experimental feasibility of a paired-atom Mott insulator with fermionic atoms, we introduce a three-component Fermi-Hubbard model (TFH model) shown in Fig. \ref{fig:theory}(a), 
\begin{align}
\hat{H}&=-\sum_{\langle ij\rangle}\sum_{s=a,b,c}t_s(\hat{s}_i^{\dagger}\hat{s}_j+H.c.) \notag\\ 
&+\sum_i\sum_{s\neq s'} U_{ss'} \hat{n}_{is}\hat{n}_{is'}.
\end{align}
Here $t_a, t_b,t_c$ are the tunneling energy of each species of atoms and $U_{ss'} (ss'=ab,ac,bc)$ are the on-site interaction between different atom pairs in each site. We assume each lattice site is filled with either an $a$ atom or a $b,c$ atom pair and the number of each species of atoms is half the number of lattice sites $N_{site}$. The parameters satisfy $t_a,t_b,t_c\ll U_{ac}-U_{bc}, U_{ab}-U_{bc},U_{ac}+U_{ab}+U_{bc}$. Thus, $|\hat{n}_{ia}=1\rangle$ and $|\hat{n}_{ib}=1, \hat{n}_{ic}=1\rangle$ can be mapped to spin up and down for the lowest-energy subspace [Fig. \ref{fig:theory}(a)], respectively. The low-energy effective model in a cubic lattice can be written as  
\begin{align}
\hat{H}_{\mathrm{eff}}=\sum_{<ij>}J_{z}\hat{S}_{iz}\hat{S}_{jz}-J_{xy}(\hat{S}_{ix}\hat{S}_{jx}+\hat{S}_{iy}\hat{S}_{jy}).
\label{eqn:H_eff} \vspace{-1mm}
\end{align}
Here $J_{z}=\sum_{s=a,b,c}\frac{2t_s^2}{U_{s2}}$ and $J_{xy}=\frac{4t_at_bt_c}{U_{a2}U_{b2}}+ \frac{4t_at_bt_c}{U_{a2}U_{c2}}+ \frac{4t_at_bt_c}{U_{b2}U_{c2}}$ with $U_{a2}=U_{ab}+U_{ac}, U_{b2}=U_{ab}-U_{bc}, U_{c2}=U_{ac}-U_{bc}$. $J_{xy}$ comes from the third-order perturbative terms and $J_{xy}\ll J_{z}$.

To make a quantitative study of the TFH model, we calculate the eigenenergy spectrum and the nearest-neighbor spin correlations $C_1^{(k)}=\sum_{i=1}^{L-1}\langle\psi_k|\hat{S}_{iz}\hat{S}_{i+1,z}|\psi_k\rangle/(L-1)$ of the lowest $N_S=\frac{L!}{N_a!(L-N_a)!}$ eigenstates of $\hat{H}$ under $U_{ab}=U_{ac}=-U_{bc}=1,t_a=t_b=t_c=1/7$. Here $|\psi_k\rangle$ is the $k$-th lowest eigenstates of $\hat{H}$. And $L=6,N_a=L/2$ are the lattice length and the number of $a$ atoms in our numerical calculation, respectively. The results show good consistency of the eigenenergy spectrum and the nearest-neighbor correlation functions of $\hat{H}_{\mathrm{eff}}$ [FIG. \ref{fig:theory}(b)]. And there is also a large energy gap $\Delta$ ($\Delta\gg J_z$) above the lowest $N_S$ eigenstates so that the partition function of the TFH model $Z=\mathrm{tr}(e^{-\hat{H}/k_BT})\approx \mathrm{tr}(e^{-\hat{H}_{\mathrm{eff}}/k_BT})$ when $k_BT$ is at the same order as $J_z$. Meanwhile, to validate the relation between a reduced spin-exchange interaction and the pairing of $b,c$ atoms, we calculate the spin correlation  $C_1^{(g)}=\sum_{i=1}^{L-1}\langle\psi_g|\hat{S}_{iz}\hat{S}_{i+1,z}|\psi_g\rangle/(L-1)$ and the pairing ratio between different species of atoms $\gamma_{ss'}=\sum_{i=1}^L\langle\psi_g|\hat{n}_{is}\hat{n}_{is'}|\psi_g\rangle/L$ ($ss'=ab,ac,bc$) of the ground state $|\psi_g\rangle$ of $\hat{H}$ at different $U_{bc}$. It can be seen that $C_1^{(g)}$ approaches to $-1/4$ while $\gamma_{bc}\rightarrow 0.5, \gamma_{ab}\rightarrow 0, \gamma_{ac}\rightarrow 0$ when $U_{bc}$ is decreased from $U_{bc}=1$ to $U_{bc}=-1$ [FIG. \ref{fig:theory}(c)], validating the reduced spin-exchange interaction is due to the pairing of $b,c$ atoms.

For possible experimental realization of the TFH model, the $a,b,c$ atoms can just be represented by the lowest three hyperfine levels of ${}^6\mathrm{Li}$. The three-body loss feature of the lowest three hyperfine levels of ${}^6\mathrm{Li}$ has been widely investigated experimentally \cite{Ottenstein2008,Huckans2009,Williams2009}. In the range of about $B=560\mathrm{G}\sim590\mathrm{G}$, the three-body recombination rate $K_3$ reaches a minimum $K_{3,\mathrm{min}}\approx10^{-25}\mathrm{cm^3/s}$ \cite{Ottenstein2008,Huckans2009}. In this magnetic range, $a_{ss'}$ is at the order of hundreds of Bohr radius and the bound states of $a,b,c$ atoms with positive $a_{ss'}$ are sufficiently deep so that the atom-dimer coupling can be safely ignored. In this way, the mixture of $a,b,c$ atoms in an optical lattice can just be described by the TFH model. If the lattice is filled as shown in FIG. \ref{fig:theory}(a), the three-body loss rate of $a,b,c$ atoms in each lattice site can be roughly estimated as $\Gamma_{loss}=K_3 \langle \hat{n}_3\rangle/a_{lat}^6$ \cite{Ottenstein2008,Eric2009}. Here $a_{lat}$ is the lattice constant and $\langle \hat{n}_3\rangle=\sum_i\langle\hat{n}_{ia} \hat{n}_{ib} \hat{n}_{ic}\rangle/N_{site}$ is the mean three-body occupation number in each lattice site. The estimated mean three-body occupation number $\bar{n}_3=N_S^{-1}\sum_{k=1}^{N_S}\langle\hat{n}_3\rangle_k$ of the lowest $N_S$ eigenstates in FIG. \ref{fig:theory}(b) is evaluated under a proper selection of parameters [FIG. \ref{fig:parameter}(b)]. It can be seen that $\bar{n}_3$ is quite small for the lowest $N_S$ eigenstates  when the on-site interaction of a three-body occupation is high.  If we make an approximation $\langle\hat{n}_3\rangle\approx 0.01,a_{lat}=532\mathrm{nm}$, the loss rate per lattice site will be $\Gamma_{loss}\approx 0.04\mathrm{Hz}$. For a lattice with $N_{site}\approx100$, the three-body loss rate is at the order of serval Hertz. Therefore, the Mott insulator of $a,b,c$ atoms [FIG. \ref{fig:theory}(a)] can be stable in a timescale of hundreds of milliseconds. For a tunneling energy at the order of hundreds of Hertz, it is long enough for experimental observation of the antiferromagnetic order under thermal equlibrium before severe atom loss.

Another concern is the magnitude of $J_z$ and $J_{xy}$ which determines the magnitude of required temperature. In the realization of the typical two-component Fermi-Hubbard model with ${}^6\mathrm{Li}$  \cite{Mazurenko2017}, $a_{ab}=210a_0,t=0.90\mathrm{kHz}, U=6.50\mathrm{kHz}$ can be achieved in a lattice of $V_{lat}=7.4E_R, a_{lat}=569\mathrm{nm}$ under a magnetic field of $B=576$G. Here $V_{lat}$ is the lattice depth and $E_R$ is the recoil energy. This yields an effective Heisenberg model $J\sum_{\langle ij\rangle}\mathbf{S}_i\cdot\mathbf{S}_j$ with $J=4t^2/U=0.50\mathrm{kHz}=24\mathrm{nK}$. Here we set the Planck constant and the Boltzmann constant to unity for simplicity hereafter. To make a proper estimate of the magnitude of experimentally accessible $J_z$ and $J_{xy}$, we assume the TFH model is realized in the same lattice configuration as in Ref. \cite{Mazurenko2017}. Accordingly,  $t_a=t_b=t_c=0.90\mathrm{kHz},U_{ss'}(B)=a_{ss'}(B)/210a_0*6.5 \mathrm{kHz}$ can be estimated if the magnetic field is near 576G. Under such a selection of parameters, $t_a,t_b,t_c\ll U_{ab}-U_{bc},U_{ac}-U_{bc},U_{ac}+U_{ab}+U_{bc}$ can be maintained in $570\mathrm{G}<B<590\mathrm{G}$ so that the mapping to $\hat{H}_{\mathrm{eff}}$ is always valid. The estimated $J_z,J_{xy}$ are shown in FIG. \ref{fig:parameter}(b). It can be seen that $J_z$ can be improved while $J_{xy}$ is largely reduced. Especially, $J_z=29\mathrm{nK},J_{xy}=6.3\mathrm{nK}$ can be achieved with $U_{ab}=6.5\mathrm{kHz},U_{ac}=4.1\mathrm{kHz},U_{bc}=-2.1\mathrm{kHz}$ at $B=576$G.

\begin{figure}[htbp]
	\includegraphics[width=\linewidth]{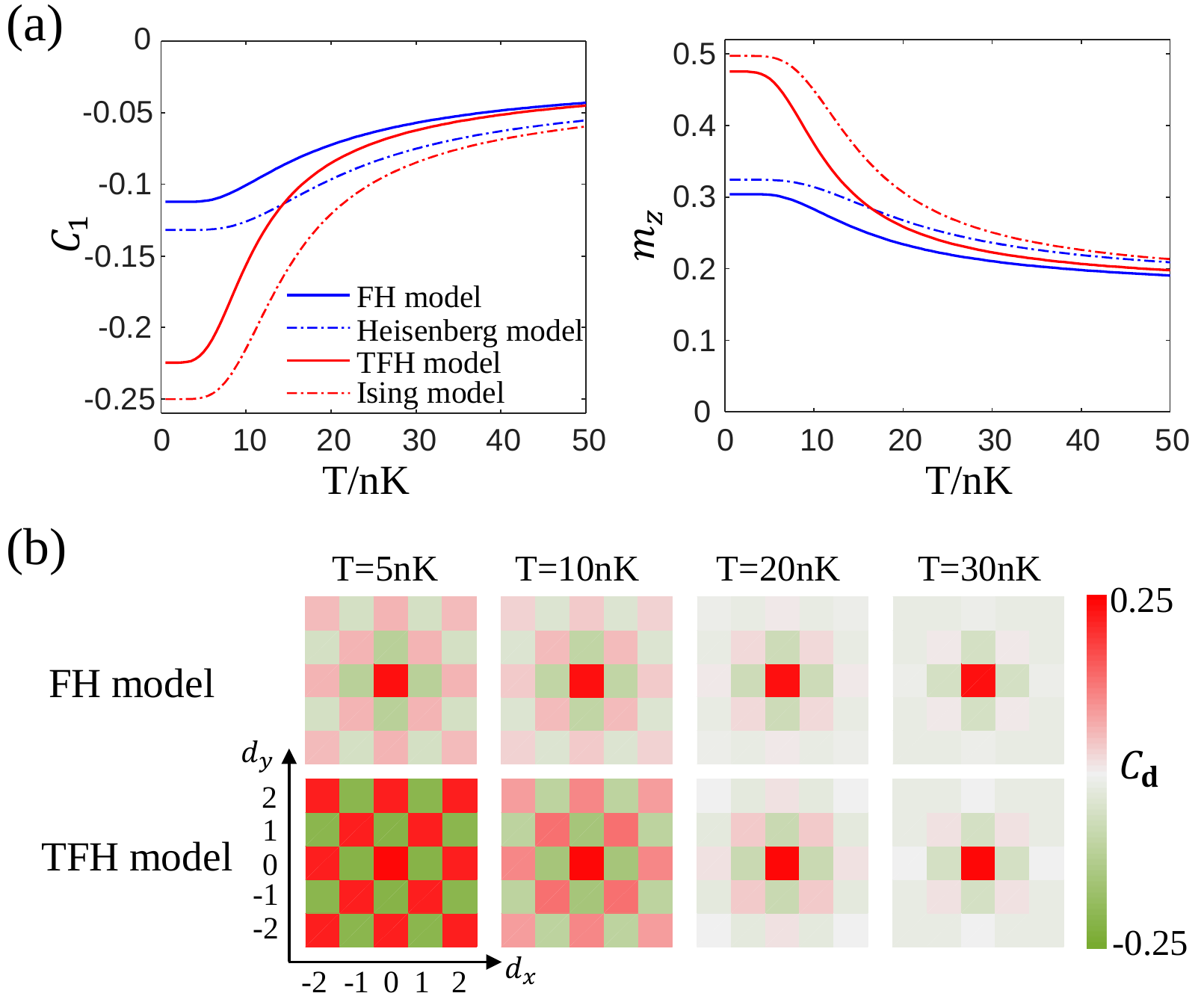}
	\caption{ Enhanced antiferromagnetism of the three-component Fermi-Hubbard model (TFH model) compared with the two-component Fermi-Hubbard  model (FH model). (a) The nearest-neighbor spin correlation $C_1$ and the staggered magnetization $m_z$ of the FH model ($J^{\mathrm{eff}}=24\mathrm{nK}$), the Heisenberg model ($J=24\mathrm{nK}$), the TFH model ($J_z^{\mathrm{eff}}=29\mathrm{nK},J_{xy}^{\mathrm{eff}}=6.3\mathrm{nK}$) and the Ising model ($J_z=29\mathrm{nK}$) in a $3\times3$ lattice. The parameters of the TFH model is assumed to be $t_a=t_b=t_c=0.90\mathrm{kHz}, U_{ss'}(B)=a_{ss'}(B)/210a_0*6.5\mathrm{kHz}$ (with $B=576$G) according to the parameters $t=0.90\mathrm{kHz},U=6.5\mathrm{kHz}$ of the FH model realized with ${}^6\mathrm{Li}$ \cite{Mazurenko2017}. (b) The long-range spin correlations  $C_{\mathbf{d}}$ of the FH model and the TFH model. It can be seen that the antiferromagnetism of the TFH model is apparently enhanced compared with the FH model when $T\leq20\mathrm{nK}$.}
	\label{fig:magnetism}
\end{figure}

To quantitatively evaluate the antiferromagnetism of the TFH model and the Fermi-Hubbard model (FH model), we calculate the effective spin correlation and the staggered magnetization 
\begin{align}
C_{\mathbf{d}}&=\frac{1}{N_{\mathbf{d}}}\sum_{\mathbf{r}}\langle\hat{S}_{\mathbf{r}}^z\hat{S}_{\mathbf{r+d}}^z\rangle, \notag\\
m_z&=\sqrt{\langle\sum_{\mathbf{r}}(-1)^{x+y}\hat{S}_{\mathbf{r}}^z/N_{site}\rangle^2}
\end{align}
of the TFH model and the FH model in a $3\times 3$ lattice. Here $\mathbf{r}=(x,y)$ and $\mathbf{d}=(d_x,d_y)$ and $N_{\mathbf{d}}$ is the number of allowed $\hat{S}_{\mathbf{r}}^z\hat{S}_{\mathbf{r+d}}^z$. From FIG. \ref{fig:magnetism}(a) and (b), it can be seen that the antiferromagnetism of the TFH model is much closer to the Ising model instead of the Heisenberg model.  And the antiferromagnetism of the TFH model is apparently enhanced in the low-temperature limit ($T\leq20\mathrm{nK}$), especially, much stronger than that of the FH model when $T$ reaches  below $5\mathrm{nK}$. 


\section{Conclusion and outlook}
In summary, we have analysed a general scheme to simulate the Ising-like magnetism  driven by superexchange interaction in a Mott insulator of paired atoms. Compared with other methods of simulating an Ising model, this method features homogeneous nearest-neighbor spin-spin interaction, broad lattice compatibiity and achiveable thermal equilibrium. And we discussed the experimental feasibility of the three-component Fermi-Hubbard model with ${}^{6}\mathrm{Li}$ atoms under a proper magnetic field where the three-body loss rate is small. Similar ideas can be applied to other mixtures of atoms. Meanwhile, it is also very promising of realizing $\hat{H}_{pair}^{BH}$ to simulate an exact classical 2D or even 3D ferromagnetic Ising model experimentally. Since the exact analytical solutions of a 2D Ising model with a $h_z$ field or a 3D Ising model is still beyond theoretical reach \cite{Barry2012}, it may promote the research of this problem by experimentally realizing an exact Ising model. Our results may be useful for experimental investigations of the classical Ising-like magnetism with ultracold neutral atoms. 


\section*{Acknowledgements}
This work is supported by the National Natural Science Foundation of China (Grants Nos. 11920101004,11334001,61727819,61475007,11825401), and the National Key Research and Development Program of China (Grant No. 2021YFA1400900, 2021YFA0718300).

%

\bibliographystyle{apsrev4-1}

\providecommand{\noopsort}[1]{}\providecommand{\singleletter}[1]{#1}%

\end{document}